\documentclass[11pt,a4paper,american]{extarticle}
\usepackage[T1]{fontenc}
\usepackage[latin1]{inputenc}
\usepackage[a4paper]{geometry}
\usepackage{amsmath}
\usepackage{amssymb}
\usepackage{setspace}
\usepackage{esint}
\usepackage{graphicx}
\usepackage{color}
\usepackage{babel}
\usepackage{bbm}
\usepackage{hyperref}
\usepackage{authblk}
\allowdisplaybreaks
\date{}

\numberwithin{equation}{section}
\author{Renann Lipinski Jusinskas\thanks{renannlj@fzu.cz}}
\affil{Institute of Physics of the Czech Academy of Sciences \\ CEICO - Central European Institute for Cosmology and Fundamental Physics
\authorcr  Na Slovance 2, 182 21, Prague - Czech Republic}

\pdfoutput=1

\begin{document}
\title{Strings as particle arrays}
\maketitle
\begin{abstract}
This work discusses the emergence of the Polyakov action from the low-energy limit of an array of relativistic particles with nearest neighbour interactions, which is suggestive of a ``microscopic'' description of string theory.
\end{abstract}

\maketitle

\section{Introduction}

Discrete descriptions of physical phenomena tend to be rewarding in
different aspects. They can be roughly divided into two classes that
often overlap. In one class, the system of interest is placed on a
lattice comprising one or more of the continous dimensions of the
original theory. This lattice model is more amenable to computational
techniques that also give us access to non-perturbative aspects of the system, e.g. lattice field theory. 
In the other class, the system of interest emerges from the continuum limit of a more fundamental
description. In this case, we may be able to explain interesting phenomena arising from the collective behaviour of microscopic degrees of freedom (e.g. Quantum Hall Effect, topological superconductors and insulators). Alternatively, it might be that a richer microscopic structure is unveiled. Such atomization is naturally connected to the quest for the most fundamental constituents of the system. 

String theory was accidentally born in this context, with a new perspective on the building blocks
of matter as tiny vibrating strings. Thus it might appear counterintuitive to look for something smaller than the string scale. But discrete descriptions within string theory also happen to be very common. Giles and Thorn introduced a lattice approach to string theory already in \cite{Giles:1977mpa}, including interactions. This was later extended to superstrings \cite{Bergman:1995wh}. Also in the light-cone, Klebanov and Susskind showed that free strings are equivalent to a lattice gauge theory without any continuum limit \cite{Klebanov:1988ba}. Further examples include cellular strings \cite{Russo:1993yw}, causal dynamical triangulation \cite{Ambjorn:2008ta}, segmented strings \cite{Vegh:2015ska}, and a variety of models within the gauge-gravity duality. More recently, a covariant description of the ``string bits'' was proposed in \cite{Mukhopadhyay:2023yfi}, based on non-local lattice derivatives with continuous time.

Predominantly, these works fit the class of discretized theories, which are guaranteed to recover string theory in the continuum limit. Alternatively, there are fewer approaches that presuppose a more fundamental description. For instance, the so-called matrix models (BFSS \cite{Banks:1996vh}, IKKT \cite{Ishibashi:1996xs}, BMN \cite{Berenstein:2002jq}, see also \cite{Ydri:2017ncg} for a more recent review), are argued to capture the strong coupling regime of string theory, being ultimately connected to M-theory. In this work, a different, more naive, proposal to an underlying string description will be discussed. The Polyakov action \cite{Deser:1976rb,Brink:1976sc,Polyakov:1981re} is shown to appear as a \textit{low-energy} limit of an array of relativistic particles with nearest neighbor interactions containing the harmonic potential. In 1967, Susskind had already noticed the connection between the Veneziano amplitude and harmonic oscillators \cite{Susskind:1969ha} (see also \cite{Susskind:1970qz}). This is regarded by many physicists as one of the origins of string theory.\footnote{I thank Urs Schreiber for bringing this to my attention.} The approach followed here is significantly different and will hopefully add to the discussion, in particular because world-sheet reparametrization is an emergent gauge freedom. This feature is further supported by the identification of the (quantum) Virasoro algebra.

\section{A simple setup}

The starting model is an array of massless particles, with action
\begin{eqnarray}
S_{0} & = & \sum_{n=1}^{N}\int d\tau_{n}\left(P_{n}\cdot\dot{X}_{n}-\frac{1}{2}e_{n}P_{n}^{2}\right),
\end{eqnarray}
where $\tau_{n}$ denotes the wordline parametrization of the $n$-th
particle, with $\dot{X}_{n}^{\mu}=dX_{n}^{\mu}/d\tau_{n}$. The worldline
metric (\emph{einbein}) is denoted by $e_{n}$, and $P_{n}^{2}=\eta^{\mu\nu}P_{n\mu}P_{n\nu}$
($\eta_{\mu\nu}$ is the Minkowski metric). The target-space coordinates
$X^{\mu}_n$ live in $D$ dimensions. Periodic boundary conditions are assumed, e.g.  $X_{n+N}^{\mu}=X_{n}^{\mu}$, and $N$ is taken to be \textit{odd}\footnote{It is interesting to note that some properties of even $N$ arrays hint at some reducibility (i.e., stable subsystems that decouple from each other), but this will not be explored in this work.}. As long as there are no
interactions between the particles, the worldline parameters are independent.
$S_{0}$ is invariant under reparametrizations of $\tau_{n}$.

The simplest nearest neighbors interaction to be considered is the harmonic potential,
\begin{equation}
V=\frac{1}{2}\sum_{n=1}^{N}\int d\tau f_{n}(X_{n+1}-X_{n})^{2},\label{eq:harmonic}
\end{equation}
with resulting action $S'=S_0-V$. Different couplings $f_{n}$ for each site $n$ are allowed. By
turning on interactions, one colapses the parameters $\tau_{n}$
to a single one, $\tau$. In order to retain an overall $\tau$-reparametrization, $f_{n}$ should be $\tau$-dependent. The fields $e_{n}$ and $f_{n}$ can be seen as Lagrange multipliers,
respectively enforcing the constraints $P_{n}^{2} = 0$ and $(X_{n+1}-X_{n})^{2} =0$. 
These constraints, however, are too stringent, implying that both the movement of individual particles and their separation from immediate neighbors are light-like.\footnote{I thank A. Mikhailov for raising this observation.}  Instead, we can replace them by a combined nearest neighbor interaction, shaped by the constraint
\begin{equation}
\mathcal{P}_n^2 +\frac{1}{\ell^4} (X_{n+1}-X_{n})^2  = 0. \label{eq:harmonic-extended}
\end{equation}
The constant $\ell$ is some characteristic length of the system, and the object $\mathcal{P}_n \equiv (P_{n+1}+P_{n})$ can be interpreted as the conjugate momentum of the mid-point $(X_{n+1}+X_{n})/2$. This equation constrains the norm of $\mathcal{P}_n$ in terms of $ (X_{n+1}-X_{n})^2$. The positive sign in \eqref{eq:harmonic-extended}  is physically meaningful, and implies a dynamical bound. A negative sign would enable arbitrary spatial separation between neighboring particles, becoming incompatible with causality.

The pair-wise interactions generated by \eqref{eq:harmonic-extended} imply a non-locality in the array, extending beyond immediate neighbors. Perhaps more surprisingly, these interactions factorize into two sectors,  ``$+$'' and ``$-$''. In order to understand how, equation \eqref{eq:harmonic-extended} can be more suggestively cast as 
\begin{equation}
(A_{n}^{+})^2 +(A_{n}^{-})^2=0,
\end{equation}
with
\begin{equation}
A_{n}^{\pm}\equiv(P_{n+1}+P_{n})\pm\frac{1}{\ell^{2}}(X_{n+1}-X_{n}),\label{eq:building-blocks}
\end{equation}
The constant $\ell$ is some characteristic length of the system, and spacetime indices were omitted for clarity.  It follows from the Poisson brackets $[X_{n}^{\mu},P_{m}^{\nu}]=\eta^{\mu\nu}\delta_{n,m}$ that
\begin{align}
[A_{n}^{\pm},A_{m}^{\pm}] =& \pm\frac{2}{\ell^{2}}(\delta_{n+1,m}-\delta_{n,m+1}),\label{eq:phi-phi-commutator} \\
[A_{n}^{+},A_{m}^{-}] =&0. 
\end{align}
The right-hand-side of the first equation showcases non-locality with respect to the array sites. Moreover, it leads to an interesting consequence when combined with the periodicity of the array, also implying that
\begin{equation}
(A_{n}^{+})^2 +(-1)^N (A_{n}^{-})^2=0. \label{eq:loc-sec}
\end{equation}
This local equation then follows from the non-locality of the interactions. Since we are working with odd $N$, the constraint \eqref{eq:loc-sec} implies the secondary constraint
\begin{equation}
\tilde{P}_n \cdot (X_{n+1}-X_{n})  = 0. \label{eq:transversal}
\end{equation}

The complete set of constraints generated by \eqref{eq:harmonic-extended} can then be summarized as
\begin{equation}
A_{n}^{+}\cdot A_{m}^{+}=A_{n}^{-}\cdot A_{m}^{-}=0.\label{eq:full-constraints}
\end{equation}
The secondary constraints involve any pair $\{n,m\}$ of sites. The message here is that turning on even a simple nearest neighbors
potential in a array of relativistic particles leads to arbitrary
pair interactions. Putting it differently, requiring \emph{local}
gauge invariance -- site-dependent transformations, consistent with
local worldline reparametrization -- leads to non-local interaction,
not confined to nearest neighbors.

The action of the resulting system may be cast as
\begin{equation}
S=\sum_{n=1}^{N}\int d\tau\bigg[P_{n}\cdot\dot{X}_{n}-\frac{1}{2}\sum_{m=1}^{N}g_{mn}^{\pm}A_{m}^{\pm}\cdot A_{n}^{\pm}\bigg].\label{eq:actiong+-h}
\end{equation}
where $g_{mn}^{\pm}$ are Lagrange multipliers. The time evolution of the system is parametrized by $\tau$, with canonical Hamiltonian given by
\begin{equation}
H_{T}=\frac{1}{2}\sum_{m,n=1}^{N}g_{mn}^{\pm}A_{m}^{\pm}\cdot A_{n}^{\pm}.
\label{eq:totalH}
\end{equation}
Reparametrization in $\tau$ is then on-shell equivalent to the gauge transformations generated by \eqref{eq:full-constraints}.

Next, it is convenient to introduce the dual array, which essentially
brings the building blocks $A_{n}^{\pm}$ to a diagonal form. It can
be implemented as
\begin{eqnarray}
X_{n}^{\mu}(\tau) & = & \ell\sum_{k=0}^{N-1}\chi_{k}^{\mu}(\tau)e^{2\pi\mathrm{i}nk/N},\\
P_{n}^{\mu}(\tau) & = & \frac{1}{N\ell}\sum_{k=0}^{N-1}\pi_{k}^{\mu}(\tau)e^{2\pi\mathrm{i}nk/N}.
\end{eqnarray}
Since $X$ and $P$ are real, their Fourier modes satisfy $\chi_{k}^{\dagger}=\chi_{N-k}$
and $\pi_{k}^{\dagger}=\pi_{N-k}$. This discrete Fourier transform
is a natural construction in this context (e.g. \cite{Bergman:1995wh}). The transition
to and from the dual array can be done using the Kronecker delta representation
\begin{equation}
\delta_{m,n}=\frac{1}{N}\sum_{k=0}^{N-1}e^{2\pi\mathrm{i}(m-n)k/N}.
\end{equation}

The mode expansion of $A_{n}^{\pm}$ can be cast as
\begin{equation}
A_{n}^{\pm}=\frac{2}{N\ell}\sum_{k=0}^{N-1}\alpha_{k}^{\pm}e^{2\pi\mathrm{i}nk/N},
\end{equation}
with
\begin{equation}
\alpha_{k}^{\pm}=\frac{(e^{2\pi\mathrm{i}k/N}+1)}{2}\pi_{k}\pm N\frac{(e^{2\pi\mathrm{i}k/N}-1)}{2}\chi_{k}.
\end{equation}
The Poisson brackets between $X_n$ and $P_n$ imply that
\begin{equation}
[\chi_{k}^{\mu},\pi_{q}^{\nu}]=\delta_{N,k+q}\eta^{\mu\nu},\label{eq:pi-chi-Poisson}
\end{equation}
which finally lead to
\begin{eqnarray}
[\alpha_{k}^{\pm},\alpha_{q}^{\pm}] & = & \pm\mathrm{i}N\sin\left(\tfrac{2\pi k}{N}\right)\delta_{N,k+q},\\{}
[\alpha_{k}^{\pm},\alpha_{q}^{\mp}] & = & 0,
\end{eqnarray}
It is worth noting that the zero mode $\chi_{0}^{\mu}$ does not
appear in $\alpha_{k}^{\pm}$, since $A_{n}^{\pm}$ is built from
$(X_{n+1}-X_{n})$. Its physical meaning is clear, denoting the collective
position of the particle array, a center-of-mass of sorts.

\section{Unleashing the dynamics}

The problem that has been so far ignored is the possible overconstraining
of the system. This can be seen by a simple
counting and comparison. We start with $N\times D$ conjugate pairs
$\{X_{n}^{\mu},P_{n}^{\mu}\}$ and, in the most general configuration,
$2\times N(N+1)/2$ constraints. Therefore, it appears that
the system becomes rigid when $N > D$, though interesting microscopic descriptions usually have unbounded $N$.

For the following discussion, it will be more suggestive to proceed in the dual array. Let us start with the $2N$ local constraints:
\begin{eqnarray}\label{eq:diagonal}
\Lambda_{n}^{\pm} & \equiv & A_{n}^{\pm}\cdot A_{n}^{\pm},\nonumber \\
 & = & \left(\frac{2}{N\ell}\right)^{2}\sum_{k=0}^{N-1}L_{k}^{\pm}e^{2\pi\mathrm{i}nk/N},\\
L_{k}^{\pm}& = & \sum_{q=0}^{N-1}(\alpha_{N-q}^{\pm}\cdot\alpha_{q+k}^{\pm}).
\end{eqnarray}

The algebra of the dual constraints $L_{k}^{\pm}$ is given by
\begin{equation}
[L_{k}^{\pm},L_{q}^{\pm}]=\pm2N(e^{-2\pi\mathrm{i}q/N}-e^{-2\pi\mathrm{i}k/N})L_{k+q,1}^{\pm},\label{eq:LL-algebra}
\end{equation}
and $L_{k,1}^{\pm}$ can be seen as a twisted version of $L_{k}^{\pm}$,
\begin{equation}
L_{k,1}^{\pm}=\sum_{q=0}^{N-1}(\alpha_{N-q}^{\pm}\cdot\alpha_{q+k}^{\pm})e^{2\pi\mathrm{i}(q+k)/N}.
\end{equation}
They correspond to the Fourier modes of $A_{n}^{\pm}\cdot A_{n+1}^{\pm}$.
More generally, we have
\begin{eqnarray}
\Lambda_{n,m}^{\pm} & = & A_{n}^{\pm}\cdot A_{n+m}^{\pm},\nonumber \\
 & = & \left(\frac{2}{N\ell}\right)^{2}\sum_{k=0}^{N-1}L_{k,m}^{\pm}e^{2\pi\mathrm{i}nk/N}\\
L_{k,m}^{\pm} & = & \sum_{q=0}^{N-1}(\alpha_{N-q}^{\pm}\cdot\alpha_{q+k}^{\pm})e^{2\pi\mathrm{i}m(q+k)/N}
\end{eqnarray}
where $L_{k,m}^{\pm}$ is the $m$-twisted version of $L_{k}^{\pm}=L_{k,0}^{\pm}$. Observe that $L_{k,N-m}^{\pm}=L_{k,m}^{\pm}$, therefore the $N^{2}+N$ constraints $\Lambda_{n,m}^{\pm}=0$ can be rewritten in the dual array as $L_{k,m}^{\pm}=0$. 

This is a good point to look at the physics of the system. We can interpret $(L^{+}_{0}+L^{-}_{0})$ as energy. This would be the Hamiltonian \eqref{eq:totalH} in the gauge $g^{\pm}_{mn}=\delta_{m,n}$. Now, note that
\begin{equation}
[L_{0}^{\pm},\alpha_{k}^{\pm}]=\mp2\mathrm{i}N\sin\left(\tfrac{2\pi k}{N}\right)\alpha_{k}^{\pm},
\end{equation}
and consider
\begin{equation}\label{eq:twisted-difference}
L_{k,1}^{\pm}-L_{k}^{\pm}=\sum_{q=1}^{N}(\alpha_{N-q}^{\pm}\cdot\alpha_{k+q}^{\pm})(e^{2\pi\mathrm{i}(q+k)/N}-1).
\end{equation}
The contribution from the lowest lying modes to the right-hand side
of this equation is of the order $\mathcal{O}(1/N)$. These modes are related to a smoothly oscillating array, in which $X_{n+1}-X_{n}\approx \ell/N$. Therefore, in a low-energy regime ($|k|,|q| \ll N$), $L_{k,1}^{\pm}$ and $L_{k}^{\pm}$ become indistinguishable as
the size of the array grows. The non-local interactions
-- involving not only nearest neighbors, described by the twisted
generators -- are suppressed when $N\gg1$. In this
case, the dynamics of the system is effectively constrained by only
$2N$ generators. For instance, equation \eqref{eq:LL-algebra} becomes
\begin{equation}
[L_{k}^{\pm},L_{q}^{\pm}]=\pm4\pi\mathrm{i}(k-q)L_{k+q}^{\pm}+\mathcal{O}(1/N).
\end{equation}
Up to order $\mathcal{O}(1/N)$, this is nothing but the classical
Virasoro algebra. The $\pm$ sign is merely a consequence of the way chosen to describe the dual array.

This analysis ensures that \eqref{eq:diagonal} does not generate non-local constraints in the low-energy limit. From the point of view of the action \eqref{eq:actiong+-h}, we are effectively left with $g_{mn}^{\pm}=\delta_{m,n} g_{m}^{\pm}$,
\begin{equation}
S=\sum_{n=1}^{N}\int d\tau(P_{n}\cdot\dot{X}_{n}-\tfrac{1}{2}g_{n}^{\pm}A_{n}^{\pm}\cdot A_{n}^{\pm})+\ldots.\label{eq:discrete-Polyakov}
\end{equation}
The ellipsis denote terms that that are suppressed with $1/N$ at low energies.

In the continuum limit, we take $N\to\infty$ while keeping a finite
$\ell$. The sum over sites is replaced by an integral over a continuos
variable $\sigma$. $P_{n}$ and $X_{n}$ are simply functions of
$\tau$ and $\sigma$ evaluated at $\sigma=n/N$. Also we have
\[
A_{n}^{\pm}\to\left.P\pm\frac{1}{\ell^{2}}\partial_{\sigma}X\right|_{\sigma=n/N},
\]
such that the continuum limit of \eqref{eq:discrete-Polyakov} is
given by the two-dimensional (world-sheet) action
\begin{equation}
\mathcal{S}=\int d\tau d\sigma\bigg[P\cdot\partial_{\tau}X-\frac{1}{2}g^{+}\bigg(P+\frac{1}{\ell^{2}}\partial_{\sigma}X\bigg)^{2}
-\frac{1}{2}g^{-}\bigg(P-\frac{1}{\ell^{2}}\partial_{\sigma}X\bigg)^{2}\bigg].\label{eq:Polyakov-1storder}
\end{equation}
This is the phase space version of the Nambu-Goto action, or the first order Polyakov action, where $g^{\pm}=g^{\pm}(\tau,\sigma)$
are Weyl-invariant Lagrange multipliers. The usual second order action is recovered after integrating out the field $P_{\mu}(\tau,\sigma)$
and identifying the two-dimensional metric components $\gamma_{ij}$
through
\begin{equation}
g^{\pm}\propto\frac{\gamma^{\tau\sigma}}{\gamma^{\tau\tau}}\mp\frac{1}{\gamma^{\tau\tau}\sqrt{-\gamma}},
\end{equation}
where $\gamma=\det\gamma_{ij}$. World-sheet reparametrization is then
an emergent gauge invariance, a feature of the low-energy limit of a
periodic array of interacting relativistic particles.

It is interesting to note that the two constraints in \eqref{eq:Polyakov-1storder} can be expressed as
\begin{equation}
P^2 + \frac{1}{\ell^2}\partial_\sigma X^2 =P\cdot \partial_\sigma X  = 0.
\end{equation} 
They are the continuous version of \eqref{eq:harmonic-extended} and \eqref{eq:transversal}. In the particle description, however, the latter is a sort of highly non-local effect that follows from the periodicity of the odd array.

\section{Some aspects of quantization}\label{sec:quantization}

The idea in this section is to take a step back and discuss an interesting
behaviour in the quantization of the system with finite $N$. The
continuum, low-energy limit would follow the traditional schemes of
bosonic string theory, yielding, for instance, a critical spacetime
dimension $D=26$.

The fields $P_{n}^{\mu}$ and $X_{n}^{\mu}$, as well as their dual modes, are promoted to operators. The Poisson brackets become equal-time commutators,\begin{subequations}
\begin{eqnarray}
[X_{n}^{\mu},P_{m}^{\nu}] & = & \mathrm{i}\hbar\eta^{\mu\nu}\delta_{n,m}.\\{}
[\chi_{k}^{\mu},\pi_{q}^{\nu}] & = & \mathrm{i}\hbar\eta^{\mu\nu}\delta_{N,k+q}.
\end{eqnarray}
\end{subequations}Because the ``$+$'' and ``$-$'' sectors
are independent, it will be enough to analyze one of them here. So let us focus
on the ``$+$'' sector and drop $\pm$ labels for clarity. 

The operator
\begin{equation}
\alpha_{k}^{\mu}=\frac{(e^{2\pi\mathrm{i}k/N}+1)}{2}\pi_{k}^{\mu}+N\frac{(e^{2\pi\mathrm{i}k/N}-1)}{2}\chi_{k}^{\mu}.
\end{equation}
satisfies
\begin{equation}
[\alpha_{k}^{\mu},\alpha_{q}^{\nu}]=\hbar N\sin\left(\tfrac{2\pi q}{N}\right)\eta^{\mu\nu}\delta_{N,k+q}.
\end{equation}
The transition from classical fields to quantum
operators has to be carefully done, because of ordering ambiguities. Products of $\alpha_{k}^{\mu}$
will be ordered according to their position in the dual array,
\begin{equation}
:\alpha_{k}^{\mu}\alpha_{q}^{\nu}:=\begin{cases}
\alpha_{k}^{\mu}\alpha_{q}^{\nu} & 0\leq q\leq k\leq N-1,\\
\alpha_{q}^{\nu}\alpha_{k}^{\mu} & N-1\geq q>k\geq0,
\end{cases}
\end{equation}
which is well defined for any array length $N$.

The ordered $m$-twisted operators are given by
\begin{equation}
L_{k,m}=\sum_{q=0}^{N-1}:(\alpha_{N-q}\cdot\alpha_{q+k}):e^{2\pi\mathrm{i}m(q+k)/N}.
\end{equation}
The ordering plays an important role in their algebra, which is straightforward
to derive. Starting with,
\begin{equation}
[L_{k,m},\alpha_{q}^{\mu}]=\hbar N\sin\left(\tfrac{2\pi q}{N}\right)
(e^{2\pi\mathrm{i}m(q+k)/N}+e^{-2\pi\mathrm{i}mqN})\alpha_{k+q}^{\mu},
\end{equation}
it is possible to show that
\begin{eqnarray}
[L_{k,m},L_{q,n}] &=&+\mathrm{i}\frac{\hbar N}{2}\left(e^{-2\pi\mathrm{i}(m+1)q/N}-e^{-2\pi\mathrm{i}(n+1)k/N}\right)L_{k+q,m+n+1}\nonumber \\
& &-\mathrm{i}\frac{\hbar N}{2}\left(e^{-2\pi\mathrm{i}(m-1)q/N}-e^{-2\pi\mathrm{i}(n-1)k/N}\right)L_{k+q,m+n-1}\nonumber \\
& &+\mathrm{i}\frac{\hbar N}{2}\left(e^{-2\pi\mathrm{i}(m-n+1)q/N}-e^{2\pi\mathrm{i}(mk+nq)/N}e^{-2\pi\mathrm{i}(n-m-1)q/N}\right)L_{k+q,m-n+1}\nonumber \\
& &-\mathrm{i}\frac{\hbar N}{2}\left(e^{-2\pi\mathrm{i}(n-m+1)k/N}-e^{2\pi\mathrm{i}(mk+nq)/N}e^{-2\pi\mathrm{i}(m-n-1)k/N}\right)L_{k+q,m-n-1}\nonumber \\
& &+(\hbar N)^{2}D\delta_{N,k+q}\left(\sum_{r=1+k}^{\lfloor N/2\rfloor+k}C_{N}(r,k,m,n)-\sum_{r=1}^{\lfloor N/2\rfloor}C_{N}(r,k,m,n)\right).\label{eq:m-twisted-algebra}
\end{eqnarray}
with
\begin{equation}
C_{N}(r,k,m,n)=\sin \left(\tfrac{2\pi r}{N} \vphantom{\tfrac{2\pi (r-k)}{N}}\right)\sin \left(\tfrac{2\pi (r-k)}{N}\right)
(e^{2\pi\mathrm{i}mk/N}e^{-2\pi\mathrm{i}(m+n)r/N}+e^{2\pi\mathrm{i}(m-n)r/N}).
\end{equation}
The last line of \eqref{eq:m-twisted-algebra} is the central charge of the algebra, a quantum effect. It receives contributions from modes around $r=0$ and $r=\lfloor N/2\rfloor$ (the floor of $N/2$), which can be seen from the rewriting of the sums ranges,
\begin{equation}\label{eq:two-regions}
\sum_{r=1+k}^{\lfloor N/2\rfloor+k}-\sum_{r=1}^{\lfloor N/2\rfloor} =\sum_{r=\lfloor N/2\rfloor+1}^{\lfloor N/2\rfloor+k}-\sum_{r=1}^{k}.
\end{equation}
This equality holds for $k>0$. For $k=0$, the left hand side is trivially zero, and the central charge vanishes. Otherwise, there is a non-vanishing central charge, due to an asymmetric input to the zero-point energies for odd $N$.

Finally, one can check the low-energy limit $|k| \ll N$ at the level of the algebra. The contribution to the central charge is proportional to
\begin{equation}
\lim_{N\to\infty}\left(\frac{N}{4\pi}\right)^{2}\left(\sum_{r=1}^{k}C_{N}(r,k,m,n)\right)
=-\frac{1}{6}k(k^{2}-1),
\end{equation}
recovering the usual Virasoro central charge for two-dimensional scalars.

\section{Outlook}

In this work the Polyakov action was shown to naturally appear as an effective description of the low-energy physics of an array of interacting massless particles. The emergence of the Virasoro algebra is an interesting feature, as there is no geometrical input, just a simplet set of nearest neighbors interactions. A natural follow-up is to investigate whether the invariance under general coordinate transformations can be seen as an emergent property in higher dimensions.

\begin{figure}[h]
\includegraphics[width=10cm]{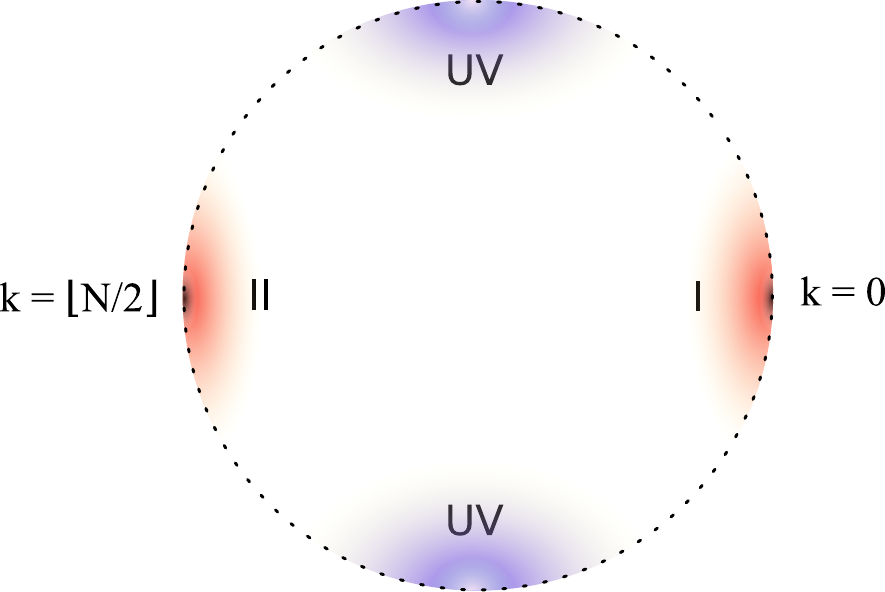}
\centering
\caption{Schematic representation of the dual array and the distribution of modes. There are two low-energy regions (I and II), around $k=0$ and $k=\lfloor N/2 \rfloor$.}
\label{fig:dual-array}
\end{figure}

The quantization of the system in section \ref{sec:quantization} should be analyzed in greater detail, in particular with respect to the physical spectrum. At the quantum level, we see there are two contributions to the central charge of the constraint algebra, cf. the decomposition \eqref{eq:two-regions}. They both come from low-energy regions of the dual array (see figure \ref{fig:dual-array}). Classically, such regions are associated to a smoothly bound array, in which $X_n$ changes slowly with $n$. They are separated by high-energy (UV) regions, which is a non-trivial ingredient in the quantum theory. In the low-energy limit discussed here ($k \ll N$), the region $k \sim \lfloor N/2 \rfloor$ becomes unaccessible, and the physical constraints reproduce the expected quantum Viraroso algebra. For the analysis of the physical spectrum, something akin to the so-called old covariant quantization in the ordinary strings would be an obvious starting point. However, the distinction between creation and annihilation operators in a periodic array is blurred, and a consistent imposition of the physical constraints is yet to be found. Alternatively, we may follow a standard  Becchi--Rouet--Stora--Tyutin (BRST) approach, introducing ghost fields and analyzing the cohomology of the related BRST charge.

There are several potentially interesting directions to explore the particle array. For instance, investigating how different two-dimensional topologies and their respective modular structure appear in this description. From the string perspective, different topologies are connected to string loops. Therefore, understanding how the particles interact in the array should be a good starting point. More ambitiously, it would be instructive (and possibly useful) to formulate a second quantized version of the particle array, aiming at a connection to string field theory.

The extension to supersymmetric models, using particle arrays with either world-line or target-space supersymmetry, should be straightforward. It would be interesting to study the emergence of the Green-Schwarz superstring from a superparticle array, in particular the role played by $\kappa$-symmetry and its world-sheet realization.

\

\textbf{Acknowledgments:} I would like to thank T. Azevedo, S. Chakrabarti, H. Gomez, C. Lopez-Arcos, and especially P. R. S. Gomes, A. Mikhailov,  and U. Schreiber for valuable suggestions and discussions. I would like to thank also the anonymous referee for useful discussions. This work was partially supported by the European Structural and Investment Funds and the Czech Ministry of Education, Youth and Sports (project FORTE CZ.02.01.01/00/22\_008/0004632), and by FAPESP (process 2024/03803-1) through a Visiting Researcher Award hosted by IFT-UNESP and ICTP-SAIFR.


\begin{thebibliography}{10}
\bibitem{Giles:1977mpa}
R.~Giles and C.~B.~Thorn,
``A Lattice Approach to String Theory,''
Phys. Rev. D \textbf{16} (1977), 366
doi:10.1103/PhysRevD.16.366

\bibitem{Bergman:1995wh}
O.~Bergman and C.~B.~Thorn,
``String bit models for superstring,''
Phys. Rev. D \textbf{52} (1995), 5980-5996
doi:10.1103/PhysRevD.52.5980
[arXiv:hep-th/9506125 [hep-th]].

\bibitem{Klebanov:1988ba}
I.~R.~Klebanov and L.~Susskind,
``Continuum Strings From Discrete Field Theories,''
Nucl. Phys. B \textbf{309} (1988), 175-187
doi:10.1016/0550-3213(88)90237-4

\bibitem{Russo:1993yw}
J.~G.~Russo,
``Discrete strings and deterministic cellular strings,''
Nucl. Phys. B \textbf{406} (1993), 107-144
doi:10.1016/0550-3213(93)90163-J
[arXiv:hep-th/9304003 [hep-th]].

\bibitem{Ambjorn:2008ta}
J.~Ambjorn, R.~Loll, Y.~Watabiki, W.~Westra and S.~Zohren,
``A String Field Theory based on Causal Dynamical Triangulations,''
JHEP \textbf{05} (2008), 032
doi:10.1088/1126-6708/2008/05/032
[arXiv:0802.0719 [hep-th]].

\bibitem{Vegh:2015ska}
D.~Vegh,
``The broken string in anti-de Sitter space,''
JHEP \textbf{02} (2018), 045
doi:10.1007/JHEP02(2018)045
[arXiv:1508.06637 [hep-th]].

\bibitem{Mukhopadhyay:2023yfi}
P.~Mukhopadhyay,
``Covariant Closed String Bits -- Classical Theory,''
[arXiv:2307.16520 [hep-th]]

\bibitem{Banks:1996vh}
T.~Banks, W.~Fischler, S.~H.~Shenker and L.~Susskind,
``M theory as a matrix model: A Conjecture,''
Phys. Rev. D \textbf{55} (1997), 5112-5128
doi:10.1103/PhysRevD.55.5112
[arXiv:hep-th/9610043 [hep-th]].

\bibitem{Ishibashi:1996xs}
N.~Ishibashi, H.~Kawai, Y.~Kitazawa and A.~Tsuchiya,
``A Large N reduced model as superstring,''
Nucl. Phys. B \textbf{498} (1997), 467-491
doi:10.1016/S0550-3213(97)00290-3
[arXiv:hep-th/9612115 [hep-th]].

\bibitem{Berenstein:2002jq}
D.~E.~Berenstein, J.~M.~Maldacena and H.~S.~Nastase,
``Strings in flat space and pp waves from N=4 superYang-Mills,''
JHEP \textbf{04} (2002), 013
doi:10.1088/1126-6708/2002/04/013
[arXiv:hep-th/0202021 [hep-th]].

\bibitem{Ydri:2017ncg}
B.~Ydri,
``Review of M(atrix)-Theory, Type IIB Matrix Model and Matrix String Theory,''
[arXiv:1708.00734 [hep-th]].

\bibitem{Deser:1976rb}
S.~Deser and B.~Zumino,
``A Complete Action for the Spinning String,''
Phys. Lett. B \textbf{65} (1976), 369-373
doi:10.1016/0370-2693(76)90245-8

\bibitem{Brink:1976sc}
L.~Brink, P.~Di Vecchia and P.~S.~Howe,
``A Locally Supersymmetric and Reparametrization Invariant Action for the Spinning String,''
Phys. Lett. B \textbf{65} (1976), 471-474
doi:10.1016/0370-2693(76)90445-7

\bibitem{Polyakov:1981re}
A.~M.~Polyakov,
``Quantum Geometry of Fermionic Strings,''
Phys. Lett. B \textbf{103} (1981), 211-213
doi:10.1016/0370-2693(81)90744-9

\bibitem{Susskind:1969ha}
L.~Susskind,
``Harmonic-oscillator analogy for the veneziano model,''
Phys. Rev. Lett. \textbf{23} (1969), 545-547
doi:10.1103/PhysRevLett.23.545

\bibitem{Susskind:1970qz}
L.~Susskind,
``Structure of hadrons implied by duality,''
Phys. Rev. D \textbf{1} (1970), 1182-1186
doi:10.1103/PhysRevD.1.1182

\end{thebibliography}
\end{document}